\documentstyle[twocolumn,aps,psfig]{revtex}

\begin{document}
\draft

\twocolumn[\hsize\textwidth\columnwidth\hsize\csname @twocolumnfalse\endcsname

\title{Hole pairing and phonon dynamics \\
in generalized 2D $t-J-$Holstein models}

\author{T. Sakai$^1$, D. Poilblanc$^{1}$ and D. J. Scalapino$^{2}$}
\address{
$^{1}$Laboratoire de Physique Quantique, Universit\'e Paul Sabatier,
31062 Toulouse, France\\
$^{2}$Department of Physics, University of California Santa Barbara, CA 93106\\
}

\date{May 96}
\maketitle 

\begin{abstract}
The formation of hole pairs in the planar t-J model is studied in 
the presence of independent {\it dynamic} vibrations of the in-plane oxygen
atoms. In-plane (breathing modes) and out-of-plane (buckling modes) 
displacements are considered. We find strong evidences 
in favor of a stabilization 
of the two hole bound pair by out-of-plane vibrations of the in-plane oxygens.
On the contrary, the breathing modes weaken the binding energy of the 
hole pair. These results are discussed in the context of the superconducting
cuprates. 
\end{abstract}

\vskip2pc]
\narrowtext

The electron-phonon interaction plays the key role 
in the conventional BCS theory of superconductivity. 
It is the source of the effective (retarded) attraction between the 
electrons and hence of the dynamical effect for pair formation. 
On the contrary, in unconventional superconductors like the 
high-$T_c$ cuprates, the driving force for superconductivity 
is commonly believed to be the strong electronic correlations. 
However, it is theoretically known that, in strongly correlated systems, 
even moderate electron-phonon interactions can have drastic consequences.
For example, it can enhance charge density wave 
(CDW) and spin density wave (SDW) instabilities 
due to polaronic self-localization effect\cite{static,static2}. 
Experimentally, the observation of some oxygen isotope effect in the 
high-$T_c$ cuprates\cite{isotope} has given evidences for
some contribution of the electron-phonon interaction in the superconductivity, 
even though the dominant pairing mechanism is due to strong 
antiferromagnetic correlations. 
The interplay between strong electronic correlations 
and electron-phonon interaction still remains an open question. 
\begin{figure}[htb]
\begin{center}
\mbox{\psfig{figure=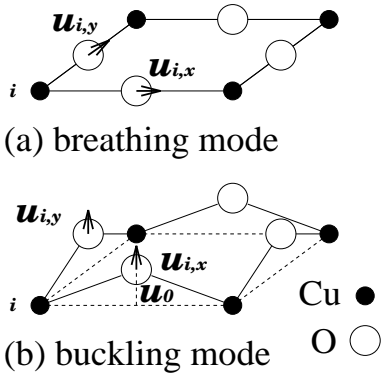,width=5cm,height=5cm,angle=-90}}
\end{center}
\caption{Schematic lattice displacements of 
breathing (a) and buckling (b) modes in the CuO$_2$ plane.
\label{fig1} }
\end{figure}

For the sake of simplicity, we describe here the low energy 
electronic degrees of freedom by a single band t-J model. We also 
restrict ourselves to the vibrations of the in-plane oxygen atoms
of the CuO$_2$ plane which have been shown to be essential.
Two types of displacement have to be considered: 
(a) in-plane breathing modes and (b) bukling modes, as shown 
schematically in Fig. \ref{fig1}. 
When the equilibrium position of the oxygen atom 
lies away from the Cu plane by $u_0$ in Fig. \ref{fig1}(b), 
the electron-phonon interaction becomes 
linear in the oxygen displacement 
perpendicular to the plane\cite{Fukuyama}, 
as it is always the case for the breathing modes. 
Such a buckling structure is realized in YBa$_2$Cu$_3$O$_{7-\delta}$. 
In the anti-adiabatic limit the two modes of interaction 
give an effective nearest-neighbor (NN) hole-hole repulsion (a) and 
attraction (b) respectively. 
In a week coupling $t$-matrix approximation which included an RPA 
antiferromagnon spin-fluctuation exchange and a phonon exchange, 
Bulut and Scalapino\cite{Bulut} found that 
the buckling mode can enhance $d_{x^2-y^2}$ pairing. 
Using an antiferromagnetic induced hole dispersion and treating the 
electron-phonon interaction at the mean-field level, 
Nazarenko and Dagotto\cite{Dagotto} 
found that the buckling mode can give rise to a $d_{x^2-y^2}$ wave 
superconducting ground state (GS).
However, both of these results involve uncontrolled approximations 
which are inadequate for treating the Hubbard and $t$-$J$ models 
in the absence of phonons. 
Thus it is of interest to carry out a numerical investigation of this 
problem. 
Our results are based on exact diagonalization studies of small 
$t$-$J$-phonon clusters. 
In agreement with the approximate results 
we find that the breathing mode suppresses the two-hole pairing
\cite{Bulut}, 
while the buckling mode stabilizes it\cite{Bulut,Dagotto}. 
However, in addition, we have examined the effect of the phonons 
on the kinetic energy, antiferromagnetic structure factor and 
hole-hole correlations, 
giving a more detailed picture of the role of dynamic lattice 
vibrations on the hole pairing. 

The hamiltonian is a generalization of the t--J-Holstein 
Hamiltonian\cite{Heff_3bands},
\begin{eqnarray}
\label{hami}
&H& = - t \sum_{<\bf i,\bf j>, \sigma}
            ( \tilde{c}_{\bf j,\sigma}^\dagger \tilde{c}_{\bf i,\sigma}
            + \tilde{c}_{\bf i,\sigma}^\dagger \tilde{c}_{\bf j,\sigma} )
   + J \sum_{<\bf i,\bf j>}
           ( {\bf S}_{\bf i} \cdot {\bf S}_{\bf j}
           - \textstyle{1 \over 4} n_{\bf i} n_{\bf j} )  
\nonumber  
\\
   &+& \sum_{\bf i,\delta}
            ( \frac{p_{\bf i,\delta}^2}{2m} 
             +\frac{1}{2}m\Omega^2 u_{\bf i,\delta}^2 )
   + g \sum_{\bf i,\delta}
            u_{\bf i,\delta}\, 
(n_{\bf i}^h\mp n_{\bf i+\delta}^h)
\end{eqnarray}
\noindent
where $\tilde{c}_{\bf i,\sigma}^\dagger$ is the usual hole creation operator,
$n_{\bf i}$ and $n_{\bf i}^h$ are the electron and hole local densities 
respectively, m is the oxygen ion mass, $\Omega$ is the phonon 
frequency and $\delta=\bf x,\bf y$ differentiates the bonds along
the x- and y-direction respectively. 
The sign -- (+) in the last term corresponds to 
the breathing (buckling) mode. 
Throughout, energies are measured in unit of the hopping integral t.
The electron-phonon g-term involves
the coupling of each copper hole with the displacements of the four
neighboring oxygens $u_{\bf i,\delta}$ and $u_{\bf i-\delta,\delta}$. 
This is clearly different from the on-site Holstein coupling  
\cite{Holstein} which has been recently used to mimic the coupling 
with the apical oxygen modes in the framework of the t--J model
\cite{Fehske}. 
Note that the displacements $u_{\bf i, \delta}$ are considered 
throughout as {\it independent} variables. 
\begin{figure}[htb]
\begin{center}
\mbox{\psfig{figure=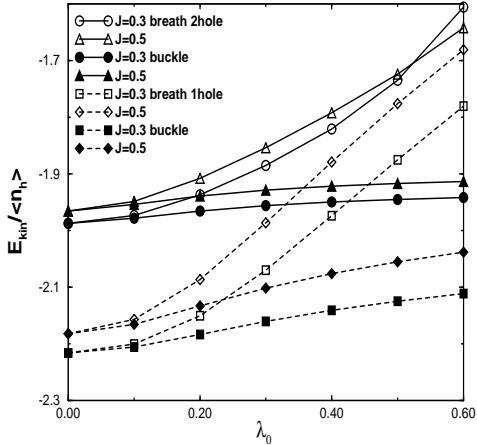,width=7cm,height=7cm,angle=-90}}
\end{center}
\caption{
Absolute value of the kinetic energy $E_{kin}$ per hole 
for $\Omega =0.2$. 
Open (Solid) symbols correspond to the breathing (buckling) mode. 
Solid (dashed) lines refer to the two (one) hole GS. 
\label{fig2} }
\end{figure}
For the purpose of our discussion it is convenient to re-write the
electron-phonon interaction in the boson representation of the phonons,
\begin{eqnarray}
\label{hamq}
H_{e-ph} =  \Omega \sum_{\bf i,\delta} 
            (&& b_{\bf i,\delta}^\dagger b_{\bf i,\delta} + \frac{1}{2}) \\
+ \lambda _0 \sum_{\bf i,\delta} 
( b_{\bf i,\delta} &&+ b_{\bf i,\delta}^{\dagger} )
(n_{\bf i}^h\mp n_{\bf i+\delta}^h)
\nonumber
\end{eqnarray}
\noindent
where 
$\lambda _0=g\,\sqrt{\frac{1}{2m\Omega}}$.
Since the phononic Hilbert space has an infinite dimension, 
we truncate it to a finite number of bosonic states 
i.e. $b_{\bf i,\delta}^{\dagger}b_{\bf i,\delta}\leq n_{ph}$ 
at each oxygen site. 
We restrict ourselves to $n_{ph}=1$. 
We have tested the validity of the one-phonon approximation 
on the $2\times 2$ 
lattice with $n_{ph}$ up to 5\cite{Didier}. 
The behaviors with the coupling constant $\lambda _0$ of the 
various relevant physical quantities are found to be unsensitive 
to $n_{ph}$. 
However, 
$n_{ph}=1$ generally {\it underestimates} the role of the phonons. 
Clearly the one-phonon calculation is a good approximation in the 
weak-coupling region ($J=0.3, \Omega =0.2, \lambda \leq 0.3$ ). 
This truncation procedure enables us to study 
$\protect\sqrt{8}\times\protect\sqrt{8}$ cluster 
with all the phonon modes (16 modes). 
We investigate the one and two-hole GS of hamiltonian 
(\ref{hami}) in a regime (0.3$\leq J \leq$ 0.5) where, 
in the absence of phonons, the two-hole pairing state is stabilized 
by the antiferromagnetic correlation, 
and we take a realistic phonon frequency $\Omega =0.2$. 
Since the $\protect\sqrt{8}\times\protect\sqrt{8}$ cluster with 
periodic boundary conditions has the $C_{4v}$ symmetry, 
we concentrate on the lowest state with the $d_{x^2-y^2}$ symmetry 
as the two-hole GS. 
Although this state is not the GS for small $J$ ($J<0.43$) 
and $\lambda _0 =0$ due to finite-size effects, 
this choice is justified 
by the fact that the two-hole GS has  
$d_{x^2-y^2}$ symmetry in the thermodynamic limit. 
\begin{figure}[htb]
\begin{center}
\mbox{\psfig{figure=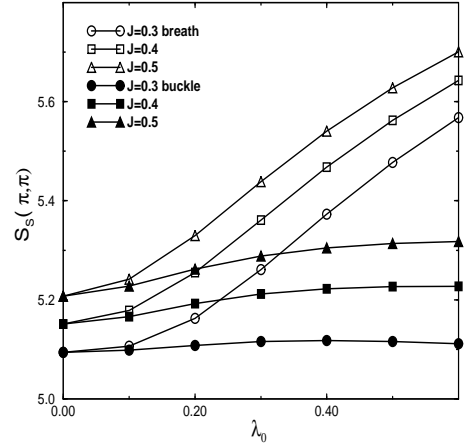,width=7cm,height=7cm,angle=-90}}
\end{center}
\caption{ Spin structure factor $S_s(\pi,\pi)$ for a single hole 
and $\Omega =0.2$. 
Open (Solid) symbols correspond to the breathing (buckling) mode. 
\label{fig3} }
\end{figure}

As a preliminary study, 
let us first briefly investigate the behavior of a single hole. 
The absolute values of the kinetic energy 
(t term in Eq. (\ref{hami})) {\it per hole} $E_{kin}$  
are shown as a function of $\lambda _0$ in Fig. \ref{fig2}. 
For the one-hole GS, 
$E_{kin}$ decreases significantly with increasing $\lambda _0$ 
for the breathing mode, 
while it does not change significantly for the buckling mode. 
This is a signature 
that only the breathing mode leads to a polaronic self-trapping 
process. 
The difference between the two modes is also clear from the behavior 
of the spin structure factor in the one-hole GS 
\begin{eqnarray}
\label{spin}
S_s(\pi,\pi)=\big< 
(\sum_{\bf i} (-1)^{(i_x+i_y)} S^z_{\bf i})^2 \big> 
\end{eqnarray}
\noindent
shown in Fig. \ref{fig3}. 
A significant increase of $S_s(\pi,\pi)$ occurs around $\lambda_0 =0.1$ 
almost independently of $J$ for the breathing mode. 
The agreement between the behaviors of 
$S_s(\pi,\pi)$ and $E_{kin}$ vs $\lambda _0$ suggests 
that the increase of the effective mass of the hole 
due to a polaronic 
self-localization effect leads, for the breathing mode, 
to an enhancement of 
the antiferromagnetic spin correlation. 
Fig. \ref{fig3} also shows that the buckling mode, 
on the contrary, does not lead to any 
crossover characteristic of self-localization. 
\begin{figure}[htb]
\begin{center}
\mbox{\psfig{figure=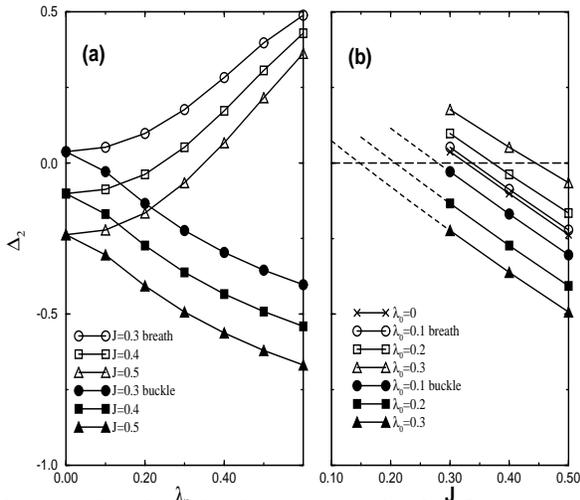,width=7cm,height=7cm,angle=-90}}
\end{center}
\caption{ Two-hole binding energy $\Delta _2$ for $\Omega =0.2$ 
(a) plotted vs $\lambda _0$ for fixed $J$, 
(b) plotted vs $J$ for fixed $\lambda $. 
Open (Solid) symbols correspond to the breathing (buckling) mode. 
\label{fig4} }
\end{figure}

The two-hole binding energy is a good probe to test the formation 
of pair of holes. 
It is defined as 
$\Delta _2 =E_0^{(2)}+E_0^{(0)}-2E_0^{(1)}$ 
where $E_0^{(n)}$ is the GS energy for a system with 
$N_h \equiv \sum _{\bf i}n^h_{\bf i}=n$. 
$E_0^{(0)}$ corresponds to 
the energy of the antiferromagnetic background. 
A negative value of $\Delta _2$ indicates the stability of a two-hole 
bound state. 
Fig. \ref{fig4}(a), where $\Delta _2$ is displayed 
as a function of $\lambda _0$, 
clearly shows that the buckling mode stabilizes 
the two-hole bound state while the breathing mode suppresses it. 
The effect of the electron-phonon interaction is to shift 
the boundary of 
the pairing phase of the $t$-$J$ model: 
the buckling mode enlarges the phase toward small $J$ while the 
breathing mode reduces it, as revealed in Fig. \ref{fig4}(b). 
The behavior of $\Delta_2$ suggests 
the possibility that the buckling mode 
assists superconductivity in the high-$T_c$ cuprates, 
while the breathing mode suppresses it. 
We note that, for the buckling mode, 
no self-trapping process occurs even in the 
two-hole state,
since there is no significant decrease of the kinetic energy 
in Fig. \ref{fig2}.
Thus the hole pair is not localized and can contribute 
to superconductivity.

\begin{figure}[htb]
\begin{center}
\mbox{\psfig{figure=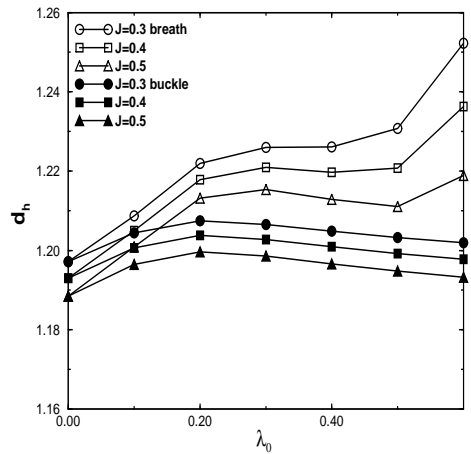,width=7cm,height=7cm,angle=-90}}
\end{center}
\caption{ Hole-hole distance in the two hole GS for $\Omega =0.2$. 
Open (Solid) symbols correspond to the breathing (buckling) mode. 
\label{fig5} }
\end{figure}

The previous data suggest that 
the electron-phonon interaction acts as an effective attraction
(repulsion) between holes for the buckling (breathing) mode, 
apparently in  agreement with the anti-adiabatic 
limit. 
However, for a finite phonon frequency, 
it is not clear, a priori, 
whether the phonon mediated interaction can be reduced to 
a static potential. 
To test this possibility, 
we consider the expectation value of the hole-hole distance 
in the two-hole GS
\begin{eqnarray}
\label{dhh}
d_h=\big< \sum_{\bf i \not= j}n^h_{\bf i}n^h_{\bf j}|{\bf j-i}|\big>
/\big< \sum_{\bf i \not= j}n^h_{\bf i}n^h_{\bf j}\big>.
\end{eqnarray}
\noindent
Fig. \ref{fig5} shows that, 
for the breathing mode, $d_h$ increases monotonously 
with increasing $\lambda _0$ 
in agreement with the effective NN hole-hole 
repulsion derived in the anti-adiabatic limit. 
However, $d_h$ for the buckling mode does not show the 
behavior expected for a NN static attraction. 
On the contrary, 
it would rather correspond to 
a small NN static repulsion, 
at least for 
small $\lambda _0 (<0.2)$. 
The failure of the anti-adiabatic picture, in this case, 
suggests that 
the effective hole-hole interaction stabilizing the hole pairing 
is controled by an essentially dynamical 
effect of the electron-phonon interaction and 
some retardation makes the range of the effective hole-hole 
attraction longer. 
In other words, the stabilization of the hole binding 
cannot be understood simply in terms of a static NN 
attraction but rather involves more subtle retardation effects. 

\begin{figure}[htb]
\begin{center}
\mbox{\psfig{figure=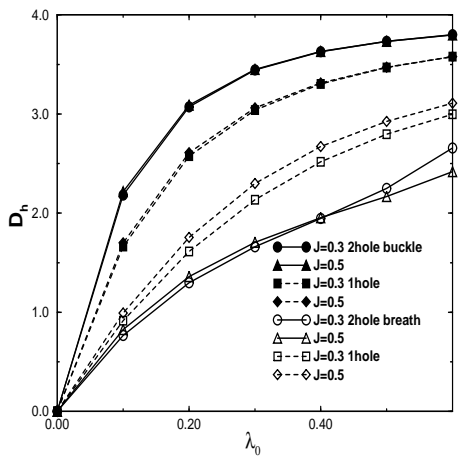,width=7cm,height=7cm,angle=-90}}
\end{center}
\caption{ Lattice deformation (per hole) around the hole sites for 
$\Omega =0.2$. 
Open (Solid) symbols correspond to the breathing (buckling) mode. 
Solid (dashed) lines refer to the two (one) hole GS. 
\label{fig6} }
\end{figure}

The dynamical effect of 
the electron-phonon interaction can be estimated 
by the lattice deformation (per hole) around the hole sites 
\begin{eqnarray}
\label{dist}
D_h=
- \big< \sum_{\bf i,\delta} 
( b_{\bf i,\delta} &&+ b_{\bf i,\delta}^{\dagger} )
(n_{\bf i}^h\mp n_{\bf i+\delta}^h)\big> /N_h,
\end{eqnarray}
\noindent
where -- (+) corresponds to the breathing (buckling) mode. 
It is proportional to the absolute value of the 
energy of the electron-phonon interaction. 
$D_h$ is always positive, which means that the oxygen ion 
deformation toward the neighboring hole sites is favored. 
$D_h$ is essentially a dynamical quantity which should 
be distinguished from the total lattice deformaton $D_{tot}$ 
given by replacing $(n_{\bf i}^h\mp n_{\bf i+\delta}^h) /N_h$ 
by unity in the form (\ref{dist}). 
The static quantity $D_{tot}$ is always zero except 
where $\Omega =0$. 
Fig. \ref{fig6} shows that, for the buckling mode, 
the deformation per hole $D_h$ in the two-hole 
state is larger than the one in the single hole state, 
in contrast to the case of the breathing mode. 
Thus, with buckling modes, 
the pair takes advantage 
of a larger deformation around the holes. 
On the contrary, the breathing deformations lead to an energy loss 
in the pairing state. 
Thus,
the relative change of the lattice contraction around the holes 
in the paired state ultimately 
contribute to a decrease or an increase of the binding 
energy. 
Particularly for the buckling mode, 
the energy gain coming from the lattice deformation clearly 
dominates the behavior of the binding energy, 
since no significant change appears in the kinetic energy 
(Fig. \ref{fig2}) or in the antiferromagnetic correlation 
which can be estimated 
from the spin structure factor $S_s(\pi,\pi)$ in Fig. \ref{fig3}. 
In other words, buckling modes stabilize the hole pairing state 
dynamically, with little changes in the static features. 

This is to be contrasted to the case of the breathing mode 
which can modify some static properties. 
As $\lambda _0$ increases, 
the effective mass becomes larger due to the polaronic self-trapping 
process and, as a result, the antiferomagnetic correlation increases. 
Since the self-localization effect in the two-hole state is smaller 
than the one-hole state (Fig. \ref{fig2}), 
this effect might tend to stabilize some trapped two-hole bound 
states. 
However, 
a larger effective repulsion due to the dynamical lattice deformation 
overcomes the static effect and suppresses the hole pairing 
in the presence of breathing modes. 

In summary, 
exact diagonalization studies of the 
generalized 2D $t$-$J$-Holstein model 
%on a $\protect\sqrt{8}\times\protect\sqrt{8}$ cluster 
give evidence for a stabilization of the two-hole pairing 
by out-of-plane buckling vibrations of the in-plane oxygens 
in the high-$T_c$ cuprates. 
On the contrary, 
in-plane breathing modes suppress the pairing. 
The difference comes from the dynamical effect of the lattice 
displacements which cannot be reduced to a simple NN static 
interaction. 
In addition, we found that the buckling mode does not give rise to 
any significant polaronic self-localization effect, 
in contrast to the breathing mode. 

{\it Laboratoire de Physique Quantique, Toulouse} is 
{\it Unit\'e Mixte de Recherche du CNRS C5626}. 
DP and DJS acknowledge support from the EEC Human Capital and Mobility 
program under Grant No. CHRX-CT93-0332 and the National Science Foundation 
under Grant No. DMR92-25027 respectively.  
We also thank IDRIS, Orsay (France), 
for allocation of CPU time on the C94 and C98 CRAY supercomputers.


\begin{references}

\bibitem{static} J. Zhong and H.-B. Sch\"uttler, Phys. Rev. Lett.
{\bf 69}, 1600 (1992).
\bibitem{static2} H. R\"oder, H. Fehske and R. N. Silver,
Europhys. Lett. {\bf 28}, 257 (1994). 
\bibitem{isotope} J. Frank, in Physical Properties
of High Temperature Superconductors IV, ed. D.M. Ginsberg,
World Scientific (1994). 
\bibitem{Fukuyama} B. Normand, H. Kohno and H. Fukuyama, 
Phys. Rev. B {\bf 53},
856 (1996).
\bibitem{Bulut} N. Bulut and D. J. Scalapino, 
Phys. Rev. B {\bf 54} (in press). 
\bibitem{Dagotto} A. Nazarenko and E. Dagotto, 
SISSA cond-mat/9510030 preprint. 
\bibitem{Heff_3bands} For derivations of 
low energy effective electron-phonon 
hamiltonians from 3-band Hubbard models see e.g.
K. von Szczepanzki and K. Becker, Z. Phys. B
{\bf 89}, 327 (1992).
\bibitem{Holstein} T. Holstein, Ann. Phys. {\bf 8}, 325 (1959).
\bibitem{Fehske} H. Fehske, H. R\"oder and G. Wellein, 
Phys. Rev. B {\bf 51}, 16582 (1995); 
G. Wellein, H. R\"oder and H. Fehske, SISSA cond-mat/9512143 preprint,
 and references therein. 
\bibitem{Didier} D. Poilblanc et al.,
Europhys. Lett. {\bf 34}, 367 (1996); T. Sakai et al., in preparation. 
\end{references}
\end{document}